\documentclass[twocolumn,aps,prc,10pt,superscriptaddress,showpacs,nofootinbib,floatfix]{revtex4-1}
\usepackage{epsfig,bm,feynmf}
\usepackage{graphicx}
\usepackage{amsmath}
\usepackage{dcolumn}
\usepackage[dvipsnames,svgnames,table]{xcolor}
 
\usepackage{hyperref}
\hypersetup{
  colorlinks,
  unicode=True,
  linkcolor=Blue,
  citecolor=Blue,
  urlcolor=Blue,
}

\usepackage{tikz-feynman,contour}
\tikzfeynmanset{compat=1.0.0}

\usepackage[normalem]{ulem}  

\renewcommand{\sout}{\bgroup \color{red} \ULdepth=-.5ex \ULset}

\begin{document}

\title{Heavy quark potential and thermal charm production in heavy-ion collisions}

\author{Taesoo Song}\email{t.song@gsi.de}
\affiliation{GSI Helmholtzzentrum f\"{u}r Schwerionenforschung GmbH, Planckstrasse 1, 64291 Darmstadt, Germany}

\author{Jiaxing Zhao}\email{jzhao@itp.uni-frankfurt.de}
\affiliation{Helmholtz Research Academy Hessen for FAIR (HFHF),GSI Helmholtz Center for Heavy Ion Research. Campus Frankfurt, 60438 Frankfurt, Germany}
\affiliation{Institut f\"ur Theoretische Physik, Johann Wolfgang Goethe-Universit\"at, Max-von-Laue-Str.\ 1, D-60438 Frankfurt am Main, Germany}

\author{Ilia Grishmanovskii}
\email{grishm@itp.uni-frankfurt.de}
\affiliation{Institut f\"ur Theoretische Physik, Johann Wolfgang Goethe-Universit\"at, Max-von-Laue-Str.\ 1, D-60438 Frankfurt am Main, Germany}

\begin{abstract}
    The effective mass of heavy quark in QGP is related to the heavy quark potential at a large distance. In this study we test different heavy quark potentials, namely, the screened potential such as the free energy and the internal energy of the heavy quark pair in QGP, and the unscreened potential, which was recently proposed by the HotQCD Collaboration, through the thermal production of charm quarks in heavy-ion collisions at the LHC. We find that the free energy potential overestimates charm production in heavy-ion collisions at the LHC, while the unscreened potential produces results closest to the experimental data from the ALICE Collaboration among the three potentials.
\end{abstract}

\maketitle

\section{Introduction}

Relativistic heavy-ion collisions produce a hot, dense partonic and/or hadronic matter. Heavy flavor particles, such as charm and bottom quarks, are among the promising probes for studying the properties of the produced matter.~\cite{Cao:2018ews,Rapp:2018qla,Xu:2018gux,Song:2020tfm,Zhao:2023nrz}. The heavy flavor has several advantages over the other probe particles. First, they are produced in the early stage of heavy-ion collisions and experience all stages of the matter's evolution. Second, their production can be reliably described by perturbative QCD (pQCD). In contrast, light particles are produced at the stage of hadronization, which occurs much later than the production of heavy quarks, and the hadronization process is a nonperturbative, soft process that depends on the model. Although light quarks are also produced before the hadronization, they repeat production and annihilation during the expansion of QGP. As for the heavy flavor, its large mass prompts to investigate the hadronization mechanism via heavy-flavor effective field theories~\cite{Braaten:1994bz}.

There are two different production processes for charm quarks. The first is the initial production of charm quark through the hard scattering of two nucleons in heavy-ion collisions. This process is calculated using the factorization formula, which separates nonperturbative and perturbative parts, where the former corresponds to the parton distribution function and the fragmentation function of heavy quark to form a hadron, while the latter corresponds to the partonic scattering cross section in pQCD.

One can estimate the number of produced heavy flavors in heavy-ion collisions by using the Glauber model and the experimental data (or pQCD calculations) on the production cross section of heavy flavor in $pp$ collisions~\cite{Song:2015sfa}. However, the parton distribution function is modified in heavy nucleus due to the (anti)shadowing effects. Since parton distribution decreases at small $x$ in heavy nucleus, the production of heavy flavor is suppressed in mid-rapidity and small $p_T$ in heavy-ion collisions at LHC energies~\cite{Eskola:2009uj,Song:2015ykw}.

Another production process for charm quarks is the thermal production. After heavy nuclei pass through each other, a hot dense matter is produced between them. If the temperature is high enough, charm quark pair can be thermally produced through partonic reactions such as quark-antiquark annihilation or two-gluon fusion. Since the mass of charm quark pair is a couple of GeV, temperature must be very high to activate the thermal production. However, even though the temperature is not extremely high, there is still a non-vanishing possibility for two energetic partons in a thermal bath to produce a charm quark pair.

Since early 90's, there have been studies on the thermal charm production in heavy-ion collisions at LHC energies~\cite{Muller:1992xn,Geiger:1993py,Levai:1994dx,Levai:1997bi,Zhang:2007yoa,Zhou:2016wbo}. Since the initial temperature realized at LHC is much higher than ever reached by other heavy-ion colliders, there has been expectation for the charm enhancement through thermal production. However, it turned out that the number of produced charm quarks are reasonably described only by the initial production including the shadowing effects~\cite{Andronic:2021erx}.

Some of us have recently studied the thermal production of charm by using the dynamical quasi-particle model (DQPM)~\cite{Song:2024hvv} which successfully describes the charm scattering in heavy-ion collisions~\cite{Song:2015sfa,Song:2015ykw} as well as the spatial diffusion coefficients of heavy quark from the lattice QCD (lQCD) calculations~\cite{Berrehrah:2014kba,Song:2019cqz}. The Feynman diagrams for charm production are simply obtained by rotating those for charm scattering~\cite{Song:2024hvv}. They found that the thermal production overestimates the experimental data on $R_{\rm AA}$ of $D$ meson at LHC, assuming charm quark has the bare mass in QCD Lagrangian, and one possible solution to this is increasing effective charm quark mass in QGP such that the thermal production is suppressed.

The effective charm quark mass around $T_c$ has also been studied in Ref.~\cite{Song:2024rjh} by matching the charm density or charm fugacity at $T_c$ by using the statistical model below $T_c$ and the quasi-particle model above $T_c$. They have found that charm number density is smoothly connected at $T_c$, as expected from the crossover phase transition, if the effective mass is around 1.8 GeV, which is consistent with the finding in Refs.~\cite{Grandchamp:2003uw,Song:2024hvv}.

On the other hand, the effective mass of heavy quark in QGP is related to heavy quark potential~\cite{Gubler:2020hft}.
Suppose a heavy quark pair is put in QGP. Ignoring kinetic energy, the total energy of the pair will be the sum of masses of heavy quark and heavy antiquark and the potential energy between them. Different from the QED potential, the QCD potential does not converge to zero at infinite distance because of the color confinement. A charm-anticharm quark pair separated from each other infinitely in vacuum will be a charm and anticharm mesons. In QGP, however, the pair will be the dressed charm and anticharm quarks which gain more mass than the bare charm quark.

The study on heavy quark potential is not presently finalized but still in progress. In early days the free energy of heavy quark pair in QGP, which is calculable in lattice QCD, was suggested as the heavy quark potential~\cite{Kaczmarek:2003ph,Burnier:2014ssa}. In this case, quarknoia melt at relatively low temperatures. On the other hand, the internal energy of a heavy quark pair can be considered as the potential~\cite{Satz:2005hx}. Since the internal energy has the contribution from the entropy density, it strongly confines quarknoium near $T_c$, and the dissociation temperatures are much higher than those from the free energy potential. One compromise was that the internal energy is more suitable for the initial stage of heavy-ion collisions, when a produced heavy quark pair does not have enough time to exchange energy with the environment, and then the free energy can be used as the potential after a sufficient time elapses. There have also been trials to linearly combine both energies as the heavy potential~\cite{Wong:2004zr,Gubler:2020hft}.


After that, it was found that a Minkowski-time Wilson loop provides the heavy quark potential in the infinite-time limit~\cite{Laine:2006ns,Brambilla:2008cx}. Since it is not directly calculable in lattice QCD, spectral functions are first extracted from the Euclidean-time Wilson loop, and then the Minkowski-time Wilson loop is reconstructed from them~\cite{Rothkopf:2011db}. However, the extraction of the spectral function has not yet been well established~\cite{Bala:2019cqu,Bala:2021fkm}.

Recently a complex heavy quark potential was extracted from the correlator of temporal Wilson lines at non-zero temperature in 2+1 flavor lattice QCD by the HotQCD Collaboration~\cite{Bazavov:2023dci}. The findings indicate that the real part of the potential is unscreening up to $T \approx 400$ MeV, which is much different from the free energy, previous lattice results, and also contrary to the widely held expectations as discussed in Ref.~\cite{Zhao:2024bia}. 

Considering the relation between the effective heavy quark mass in QGP and the heavy quark potential at a long distance, the thermal production of charm quark in heavy-ion collisions will provide useful information about the potential. In this study, we examine the effects of three different heavy quark potentials, the free energy, internal energy and the unscreened potential, on the thermal production of charm and compare with the experimental data in heavy-ion collisions at LHC.

This paper is organized as follows. In Sec.~\ref{sec.rate} we briefly review the thermal production of charm quark in QGP and discuss about its dependence on effective charm quark mass. Then the relation between heavy quark potential and heavy quark mass in QGP is presented. In Sec.~\ref{hic} the thermal production is implemented in heavy-ion collisions within the parton-hadron-string dynamics (PHSD), and the results are presented. Finally, a summary is given in Sec.~\ref{conclusion}.

\section{Thermal production rate}
\label{sec.rate}

A partonic matter produced in relativistic heavy-ion collisions is not a free gas but a strongly interacting matter like fluid, which is supported by strong elliptic flows measured in semi-central collisions~\cite{STAR:2000ekf,PHENIX:2003qra,ALICE:2010suc}.

The dynamical quasi-particle model (DQPM) deals with a strongly interacting partonic matter~\cite{Berrehrah:2016vzw,Moreau:2019vhw,Soloveva:2019xph}. In the DQPM (anti)quark and gluon gain thermal mass in the spectral form with a pole mass and spectral width, which originate from (anti)quark and gluon complex self-energies in QGP and, as a result, depend on temperature and baryon chemical potential. The complex self-energy is given by the hard thermal loop calculations, but the strong coupling in it is parameterized such that the equation of state from the lattice QCD is reproduced not only at zero but also non-zero chemical potentials~\cite{Moreau:2019vhw,Soloveva:2019xph}.

Heavy quark interacts in the DQPM with the massive off-shell quark and gluon. Since the exchanged quark and gluon are also massive and off-shell, there is no need to introduce a screening mass to prevent singularity and the scattering is less forward, compared with the scattering with massless partons~\cite{Berrehrah:2013mua,Berrehrah:2014kba}. The DQPM reproduces the spatial diffusion coefficients of heavy quark from lattice QCD~\cite{Banerjee:2011ra,Song:2019cqz} and also $R_{\rm AA}$ of $D$ mesons in relativistic heavy-ion collisions within the PHSD~\cite{Song:2015sfa,Song:2015ykw}.

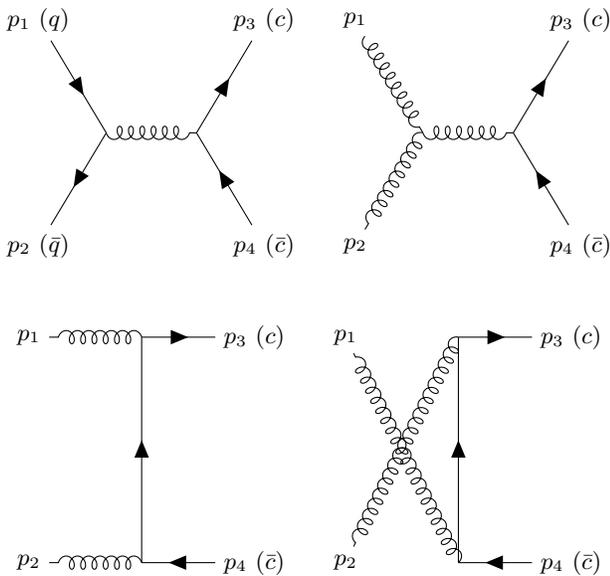
\begin{figure}[ht!]
\centerline{
    \begin{tikzpicture}
    \begin{feynman}
    \vertex (b) at (-0.6,0.);
    \vertex (a) at (-1.5,1.5) {$p_1$ ($q$)};
    \vertex (c) at (-1.5,-1.5) {$p_2$ ($\bar{q}$)};
    \vertex (e) at (0.6,0.);
    \vertex (f) at (1.5,1.5) {$p_3$ ($c$)};
    \vertex (d) at (1.5,-1.5) {$p_4$ ($\bar{c}$)};
    \diagram* {
        (a) -- [fermion] (b) -- [fermion] (c),
        (d) -- [fermion] (e) -- [fermion] (f),
        (b) -- [gluon] (e),
       };
    \end{feynman}
    \end{tikzpicture}
\quad
  \begin{tikzpicture}
    \begin{feynman}
      \vertex (b) at (-0.6,0.);
      \vertex (a) at (-1.5,1.5) {$p_1$ };
      \vertex (c) at (-1.5,-1.5) {$p_2$ };
      \vertex (e) at (0.6,0.);
      \vertex (f) at (1.5,1.5) {$p_3$ ($c$)};
      \vertex (d) at (1.5,-1.5) {$p_4$ ($\bar{c}$)};
      \diagram* {
        (a) -- [gluon] (b) -- [gluon] (c),
        (d) -- [fermion] (e) -- [fermion] (f),
        (b) -- [gluon] (e),
       };
    \end{feynman}
  \end{tikzpicture}
}
\vspace{2em}
\centerline{
  \begin{tikzpicture}
    \begin{feynman}
      \vertex (b) at (0.,1.5);
      \vertex (a) at (-1.5,1.5) {$p_1$};
      \vertex (c) at (-1.5,-1.5) {$p_2$};
      \vertex (e) at (0.,-1.5);
      \vertex (f) at (1.5,1.5) {$p_3$ ($c$)};
      \vertex (d) at (1.5,-1.5) {$p_4$ ($\bar{c}$)};
      \diagram* {
        (d) -- [fermion] (e) -- [fermion] (b) -- [fermion] (f),
        (b) -- [gluon] (a),
        (e) -- [gluon] (c),        
       };
    \end{feynman}
  \end{tikzpicture}
\quad
  \begin{tikzpicture}
    \begin{feynman}
      \vertex (b) at (0.,1.5);
      \vertex (a) at (-1.5,1.5) {$p_1$};
      \vertex (c) at (-1.5,-1.5) {$p_2$};
      \vertex (e) at (0.,-1.5);
      \vertex (f) at (1.5,1.5) {$p_3$ ($c$)};
      \vertex (d) at (1.5,-1.5) {$p_4$ ($\bar{c}$)};
      \diagram* {
        (d) -- [fermion] (e) -- [fermion] (b) -- [fermion] (f),
        (b) -- [gluon] (c),
        (e) -- [gluon] (a),        
       };
    \end{feynman}
  \end{tikzpicture}
}
  \caption{Charm quark pair production through the annihilation of a quark and antiquark pair and through the fusion of two gluons.}
\label{feynman-fig}
\end{figure}

Based on this success, some of us have studied the thermal production of charm quark in QGP. The relevant Feynman diagrams are shown in Fig.~\ref{feynman-fig}~\cite{Song:2024hvv}. The upper left diagram is for the annihilation of quark-antiquark pair and the other three for the fusion of two gluons. We note that they are nothing but the rotation of Feynman diagrams for charm quark scattering off light (anti)quark and gluon, respectively.

From the diagrams one can calculate the scattering cross sections for charm production and, simply assuming the thermalization of QGP, the production rate of charm which is defined as
\begin{eqnarray}
    \Gamma=\Gamma_{q\bar{q}\rightarrow c\bar{c}}+\Gamma_{gg\rightarrow c\bar{c}},
\end{eqnarray}
where
\begin{eqnarray}
\Gamma_{q\bar{q}\rightarrow c\bar{c}}= \sum_{q=u,d,s}\int dm_1 \rho_q(m_1)\int dm_2 \rho_{\bar{q}}(m_2)\nonumber\\
\times \int \frac{d^3k_1 d^3k_2}{(2\pi)^6 } f_q(k_1,m_1) f_{\bar{q}}(k_2,m_2)v_{q\bar{q}}\sigma_{q\bar{q}\rightarrow c\bar{c}},\\
\Gamma_{gg\rightarrow c\bar{c}}=\int dm_1 \rho_g(m_1)\int dm_2 \rho_g(m_2)~~~~~~~\nonumber\\
\times \int \frac{d^3k_1 d^3k_2}{(2\pi)^6 } f_g(k_1,m_1) f_g(k_2,m_2)v_{gg}\sigma_{gg\rightarrow c\bar{c}},
\label{rate}
\end{eqnarray}
with $\rho_i$, $f_i$, $v_{ij}$, and $\sigma_{ij}$ being the spectral function and thermal distribution function of parton $i$, the relative velocity of partons $i$ and $j$, and their scattering cross section for charm production, respectively.

\begin{figure}[ht!]
    \includegraphics[width=8.5 cm]{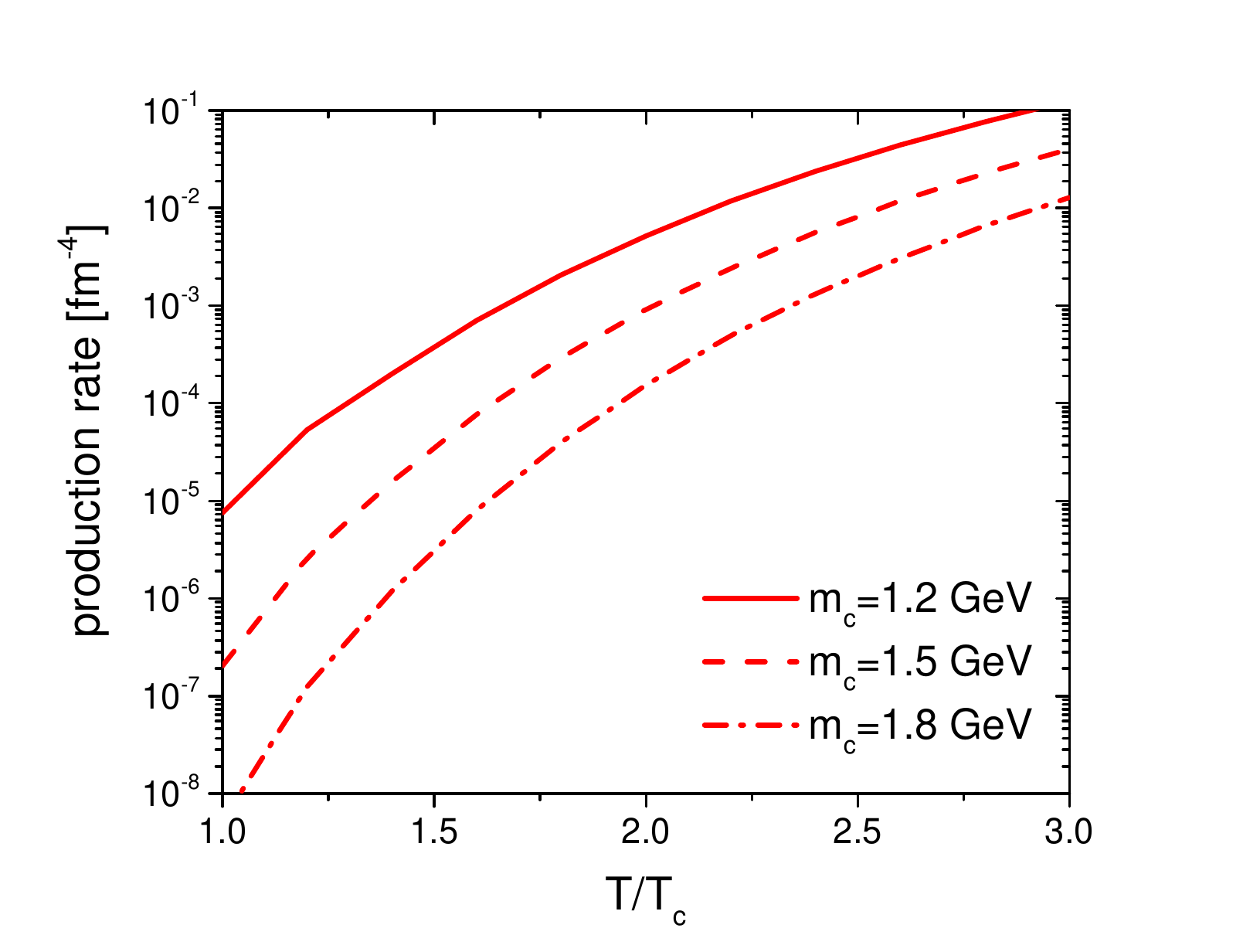}
    \caption{Thermal production rates of charm quark as a function of the temperature for three different charm quark masses in the DQPM.}
    \label{rate-mass}
\end{figure}

Figure~\ref{rate-mass} shows the thermal production rate of charm as a function of temperature for three different effective charm quark masses. As seen in the figure, the production rate is sensitive to the charm quark mass. For example, if the effective charm quark mass changes from 1.2 GeV to 1.8 GeV, the production rate decreases by three order of magnitude at $T_c$ and 10 times at $3 T_c$, even though the mass difference between them is just 0.6 GeV. The reason is that the temperatures are not high enough to produce charm quark pair. The difference will be smaller with increasing temperature.

Let us briefly explain why we call $m_c$ in Fig.~\ref{rate-mass} the effective charm quark mass.
Either in a vacuum or in the thermal medium (below $T_c$), the heavy flavor production can be treated by the QCD factorization theory, where the perturbative calculation for the $c\bar{c}$ production is carried out with almost the bare quark mass and the non-perturbative part describes the charm quark to charmed hadrons due to the absence of free quarks in the vacuum. However, this QCD factorization theory fails to describe the low energy scattering. In a vacuum, this low energy scattering is governed by the effective field theories in which the degrees of freedom are charmed hadrons such as the $D$ meson instead of the charm quark. 

For example, the threshold energy for charm production in $e^+ e^-$ scattering is not twice charm quark mass, but twice the $D$ meson mass, because two charm quarks must hadronize.
If the scattering energy of $e^+ e^-$ is above twice charm quark mass but below twice $D$ meson mass, a produced charm quark pair cannot overcome the string tension which is related to the long-range heavy quark potential and will eventually be annihilated or form a bound state.
In this respect the survived charm production is related to the long-range heavy-quark potential, though charm-quark production itself is a hard process occurring on a very short space-time scale.

In this study we focused on the $c\bar c$ thermal production in the QGP. 
In contrast to the vacuum case, the thermally produced charm quarks do not need to be confined in colorless charm hadrons in the QGP.
The main question is then how to define the effective charm quark mass in the QGP. 
One might think that the threshold energy is simply twice the charm-quark bare mass in QGP. Then charm quarks are produced more easily. This concept is analogous to the strangeness enhancement, which was proposed as a signature of QGP formation in early days~\cite{Rafelski:1982pu}. 
However, it is not the case for charm production in heavy-ion collisions at the LHC, where
the experimental data on $R_{\rm AA}$ of $D$ mesons in Pb+Pb collisions at $\sqrt{s_{\rm NN}}=5.02$ TeV favors little thermal production of charm~\cite{Song:2024hvv}.
If the threshold energy were truly twice the bare mass of the charm quark, significantly more charm quarks should have been observed in Pb+Pb collisions at the LHC.

It is well known from thermal field theory that a parton gains thermal mass in the QGP, and the concept of a massive parton is supported by on-shell and off-shell quasi-particle models that reproduce the equation of state of the QGP from lQCD~\cite{Moreau:2019vhw,Soloveva:2019xph}. Our recent study on the effective charm quark mass near $T_c$ is also consistent with it~\cite{Song:2024rjh}.

In QGP the separated heavy quark will not form $D$ mesons but two dressed charm (anti)quarks. In this respect, one can assume
\begin{eqnarray}
    m(T) =m_0+\frac{1}{2} \lim_{r\rightarrow \infty} V(r,T),
    \label{dmass}
\end{eqnarray}
where $m_0=1.26$ GeV is charm quark bare mass with which the Schr\"odinger equation can reproduce charmonium resonances in vacuum~\cite{Satz:2005hx,Zhao:2020jqu,Andronic:2024oxz}. We note that 1.26 GeV is very close to the charm quark mass in the Particle Data Group~\cite{ParticleDataGroup:2024cfk}.
In vacuum Eq.~(\ref{dmass}) is not charm quark mass but $D$ meson mass.
In this respect the effective charm quark mass in this study means a half threshold energy for charm production whether in vacuum or in QGP.

Considering that thermal production rate of charm quark is sensitive to its effective mass, one can test heavy quark potential through the thermal production in heavy-ion collisions.
Since our calculations need long range potential, it does not matter whether the produced $c\bar c$ is the color-singlet or color-octet state, because the long range potential is almost the same~\cite{Nakamura:2004wra}. Furthermore, we calculate the cross sections for charm production by using Feynman diagrams which assume long range separation of final state particles (final state particles are represented by plane waves).

There have been several models for the heavy quark potentials. For example, the free energy of heavy quark pair was suggested as the the heavy quark potential~\cite{Kaczmarek:2003ph,Kaczmarek:2005ui}. The internal energy which has additional contribution from entropy density is another candidate for the potential~\cite{Kaczmarek:2005uv}. In the former case the heavy quark potential is relatively weak and quarkonia melt at relatively low temperatures, while the potential from the internal energy is much stronger, especially near $T_c$, and quarkonia survive up to high temperatures~\cite{Satz:2005hx}. The combination of both was also suggested as the heavy quark potential~\cite{Wong:2004zr,Gubler:2020hft}. Recently a complex heavy quark potential was extracted from the correlator of temporal Wilson lines at finite temperature in lattice QCD~\cite{Bazavov:2023dci}. Surprisingly the real part of the heavy quark potential barely changes with increasing temperature, which means no color screening in QGP, and only the imaginary part of potential enhances, though it depends on how to extract the spectral function from the Euclidean-time Wilson loop~\cite{Bala:2019cqu,Bala:2021fkm}. 

In this study we test these three different heavy quark potentials by using the thermal production of charm quark in heavy-ion collisions at LHC energy. We note that this study does not confirm the heavy quark potential in all spatial range but only in the range of long distance.

\begin{figure}[ht!]
    \includegraphics[width=8.5 cm]{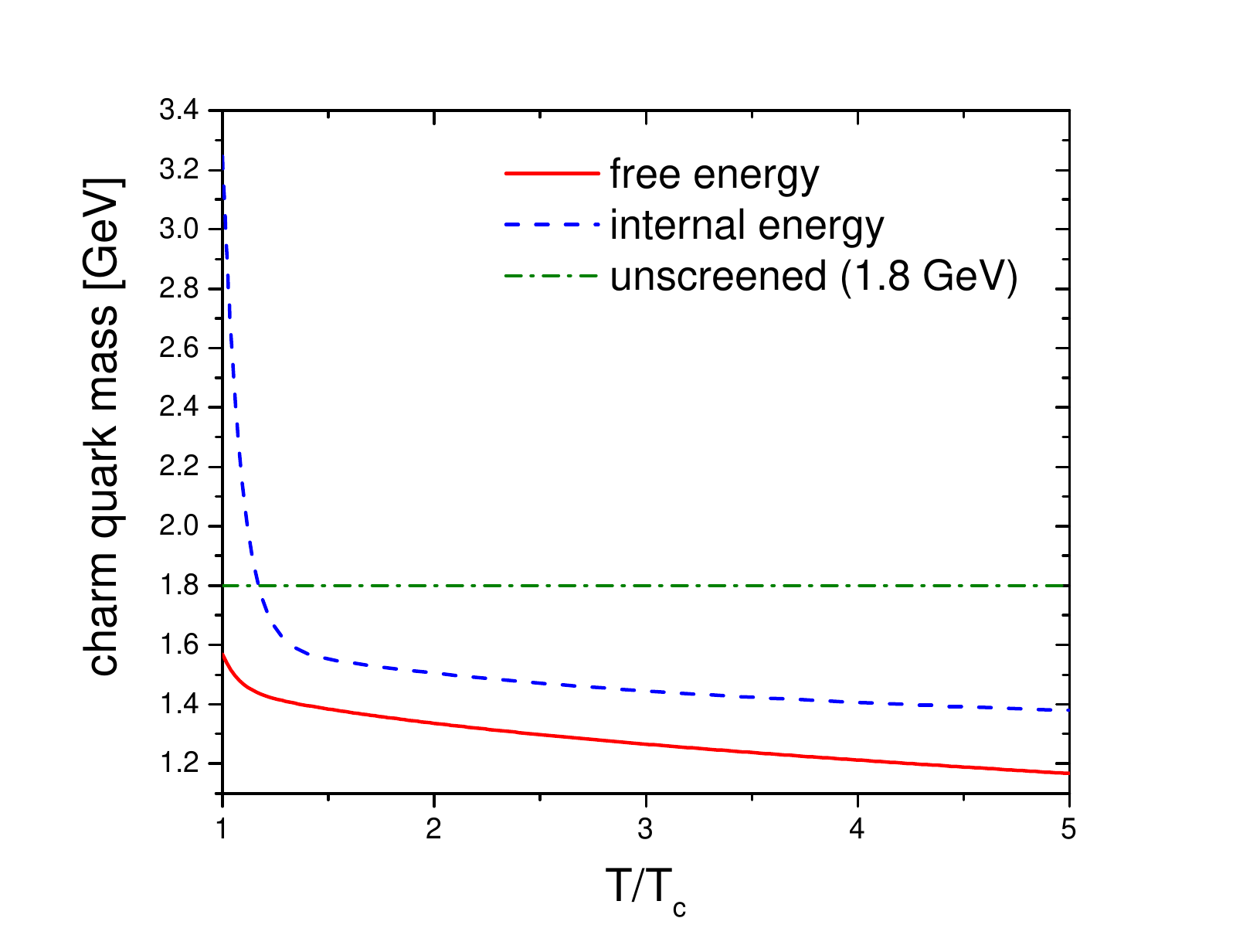}
    \includegraphics[width=8.5 cm]{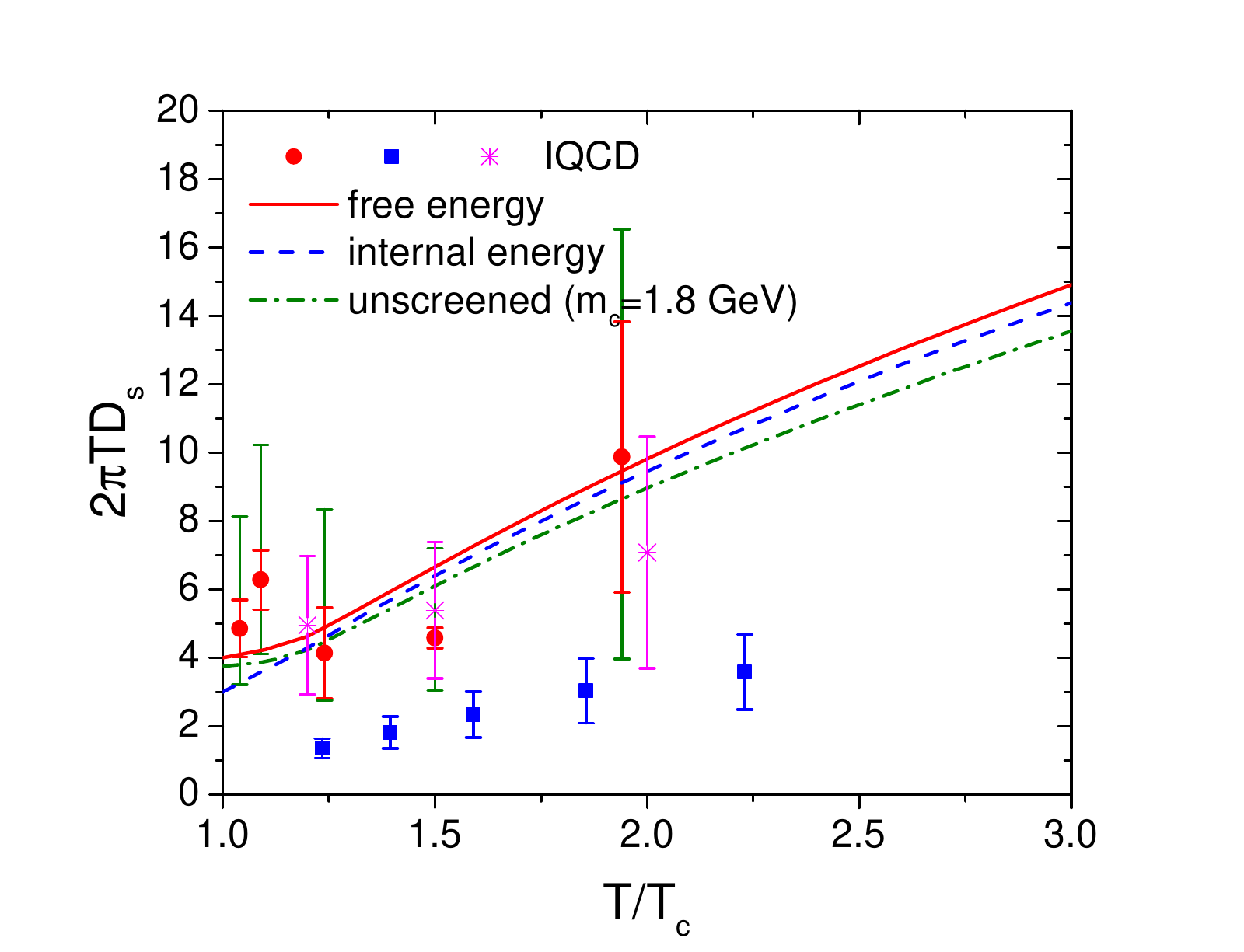}
    \caption{(Upper) effective charm quark mass as a function of temperature from the free energy and internal energy in comparison with 1.8 GeV  and (lower) the corresponding spatial diffusion coefficients $D_s$ in comparison with the lattice QCD calculations~\cite{Banerjee:2011ra,Banerjee:2022uge, Altenkort:2023eav}.}
    \label{charm-mass}
\end{figure}

Figure~\ref{charm-mass} shows effective charm quark mass as a function of temperature in three different heavy quark potentials. We assume that the effective charm quark mass from Ref.~\cite{Bazavov:2023dci} is constantly 1.8 GeV based on the recent study from the statistical model and the quasi-particle model~\cite{Song:2024rjh}. One may take $D$ meson mass for the effective charm quark mass in the third scenario, but it does not change our conclusion qualitatively, as shown in the next section. One finds that the effective charm quark mass from the internal energy is very large near $T_c$ because the entropy density $(\partial F/\partial T)$ rapidly changes during the phase transition. It is still larger above $T_c$ than that from the free energy due to the non-vanishing contribution from the entropy density. Both free energy and internal energy potentials predict the decreasing effective charm quark masses with increasing temperature, while the recent lattice QCD study supports a constant effective charm quark mass~\cite{Bazavov:2023dci}. However, the spatial diffusion coefficients of charm quark from the three scenarios are not much different from each other and all consistent with the lattice calculations~\cite{Banerjee:2011ra, Altenkort:2023eav}, as shown in the lower panel of Fig.~\ref{charm-mass}.

\section{Thermal charm production in heavy-ion collisions}
\label{hic}

In this section we apply the processes in Fig.~\ref{feynman-fig} to heavy-ion collisions by using the PHSD, which is a non-equilibrium microscopic transport approach for the description of strongly interacting hadronic as well as partonic matter produced in heavy-ion collisions~\cite{Cassing:2008sv,Cassing:2009vt}. Since the PHSD deals with non-equilibrium partonic matter microscopically, we do not need to assume thermalization or use the thermal production rate in Fig.~\ref{rate-mass}, but directly use the scattering cross sections for charm production in QGP from Fig.~\ref{feynman-fig}.

\begin{figure}[ht!]
    \includegraphics[width=8.5 cm]{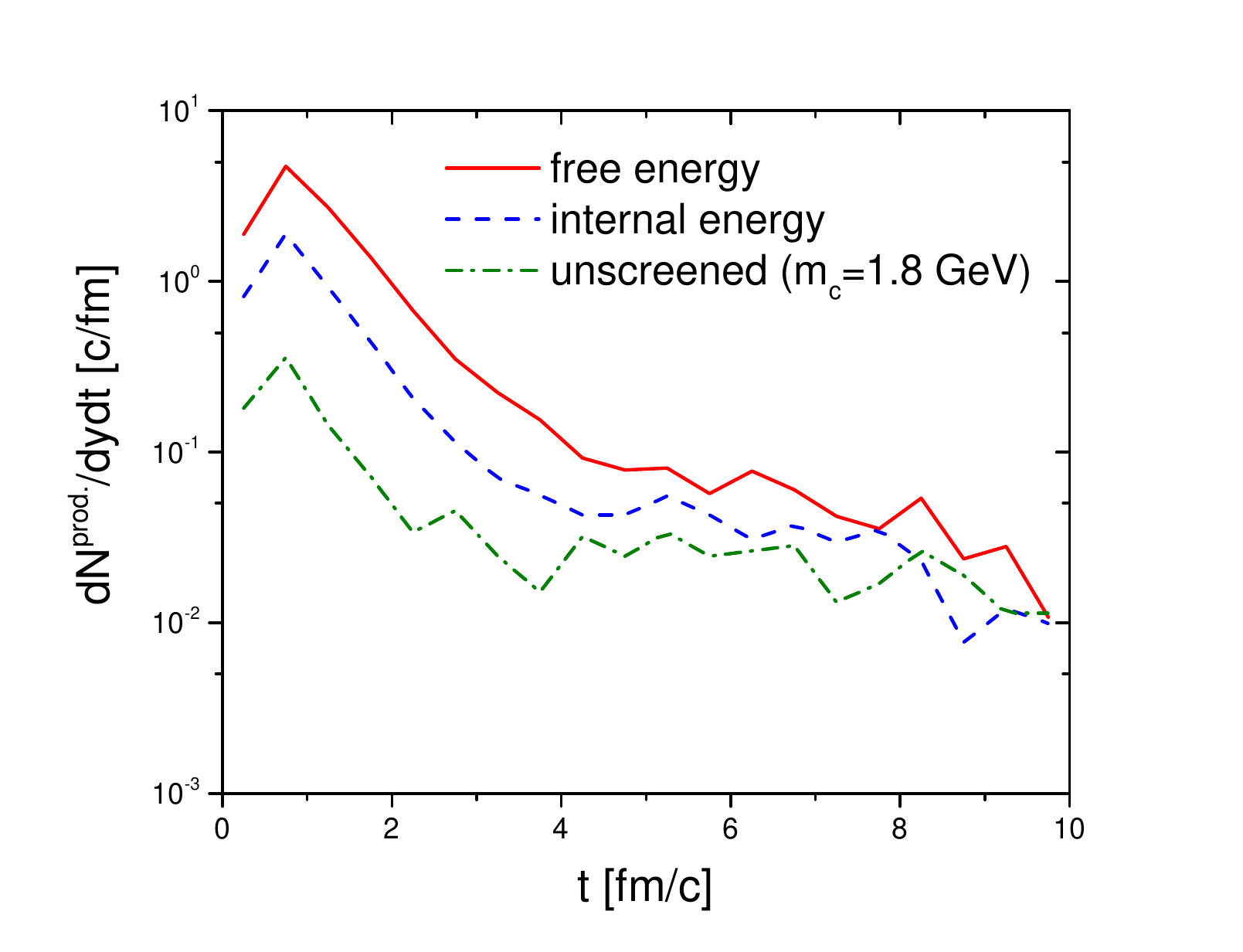}
    \includegraphics[width=8.5 cm]{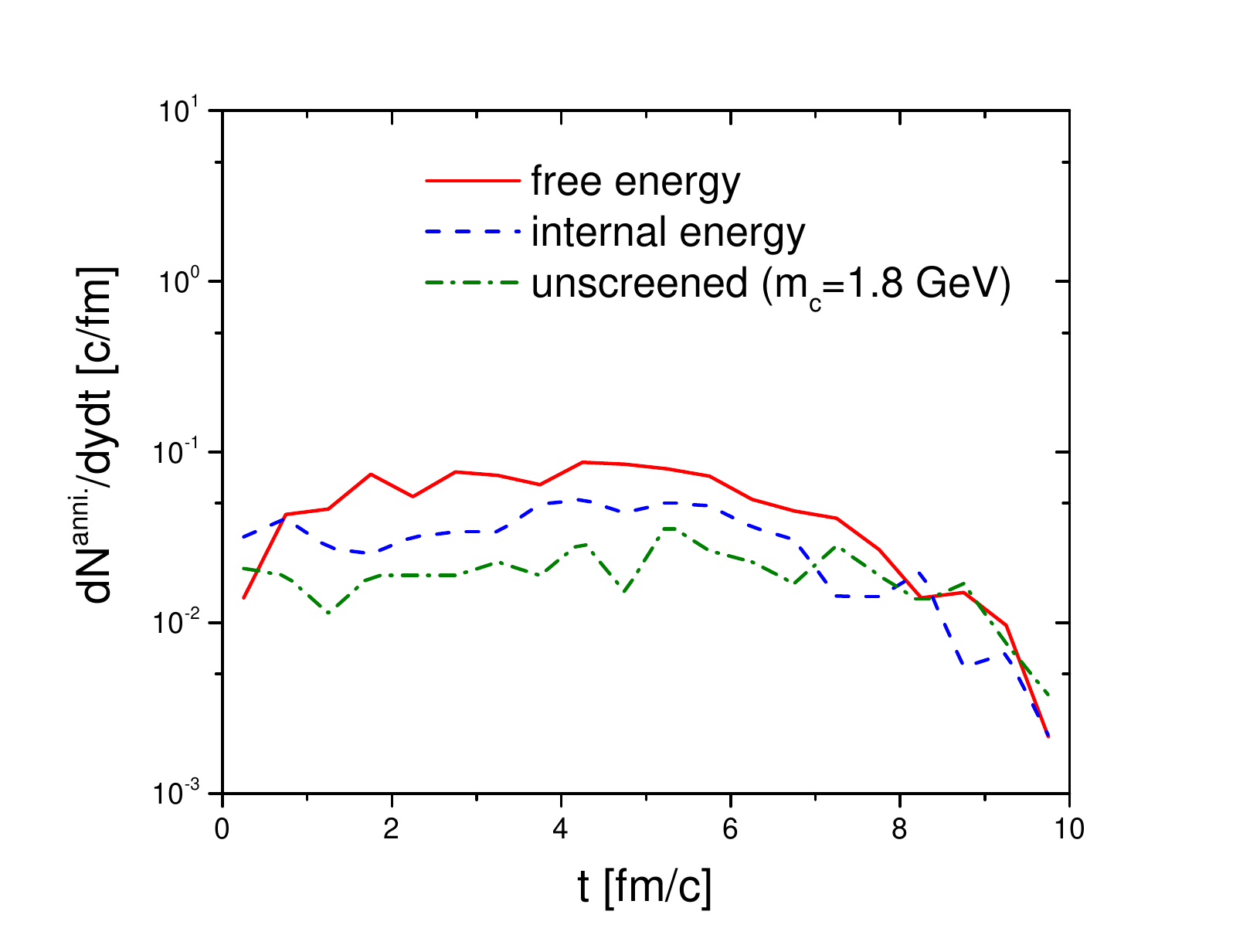}
    \caption{Thermal production and annihilation of charm quark for three different effective charm quark masses in 0-10 \% central Pb+Pb collisions at $\sqrt{s_{\rm NN}}=$ 5.02 TeV.}
    \label{prod-anni}
\end{figure}

We display in Fig.~\ref{prod-anni} the distribution of charm thermal production and annihilation for effective charm quark mass from the three different heavy quark potentials. Since each parton has a formation time, which is inversely proportional to its energy, thermal production is suppressed at the very early stage in spite of the largest local energy density. The numbers of produced charm quark pairs in mid-rapidity are respectively 6.4, 2.4 and 0.6 for effective charm quark masses from the free energy, internal energy, and unscreened potentials. On the other hand, the numbers of annihilated charm quark pairs are 0.5, 0.3 and 0.2 on average. Since the free energy potential produces more charm quarks, more annihilation happens due to the larger charm quark number or charm quark density, though the differences are little. Assuming that $d\sigma_{c\bar{c}}^{NN}/dy=1.165$ mb at $\sqrt{s_{\rm NN}}=$ 5.02 TeV~\cite{ALICE:2019nxm,ALICE:2019nrq,ALICE:2021dhb} and average unclear overlap function, $T_{\rm AA}$, in 0-10 \% central collisions is 23.3 $\rm mb^{-1}$~\cite{ALICE:2019nrq}, the expected number of charm pairs produced from the initial hard scattering is roughly 27.1 per rapidity. Simply counting, the thermal production and annihilation will enhance charm by 23 \%, 9 \% and 2 \% for effective charm quark masses from the free energy, internal energy and unscreened heavy quark potentials. However, considering the shadowing effects, which suppress initial charm production from nucleon-nucleon hard scattering~\cite{Song:2015ykw,Eskola:2009uj}, the relative contribution from the thermal production will be larger.
\begin{figure}
\includegraphics[width=8.5 cm]{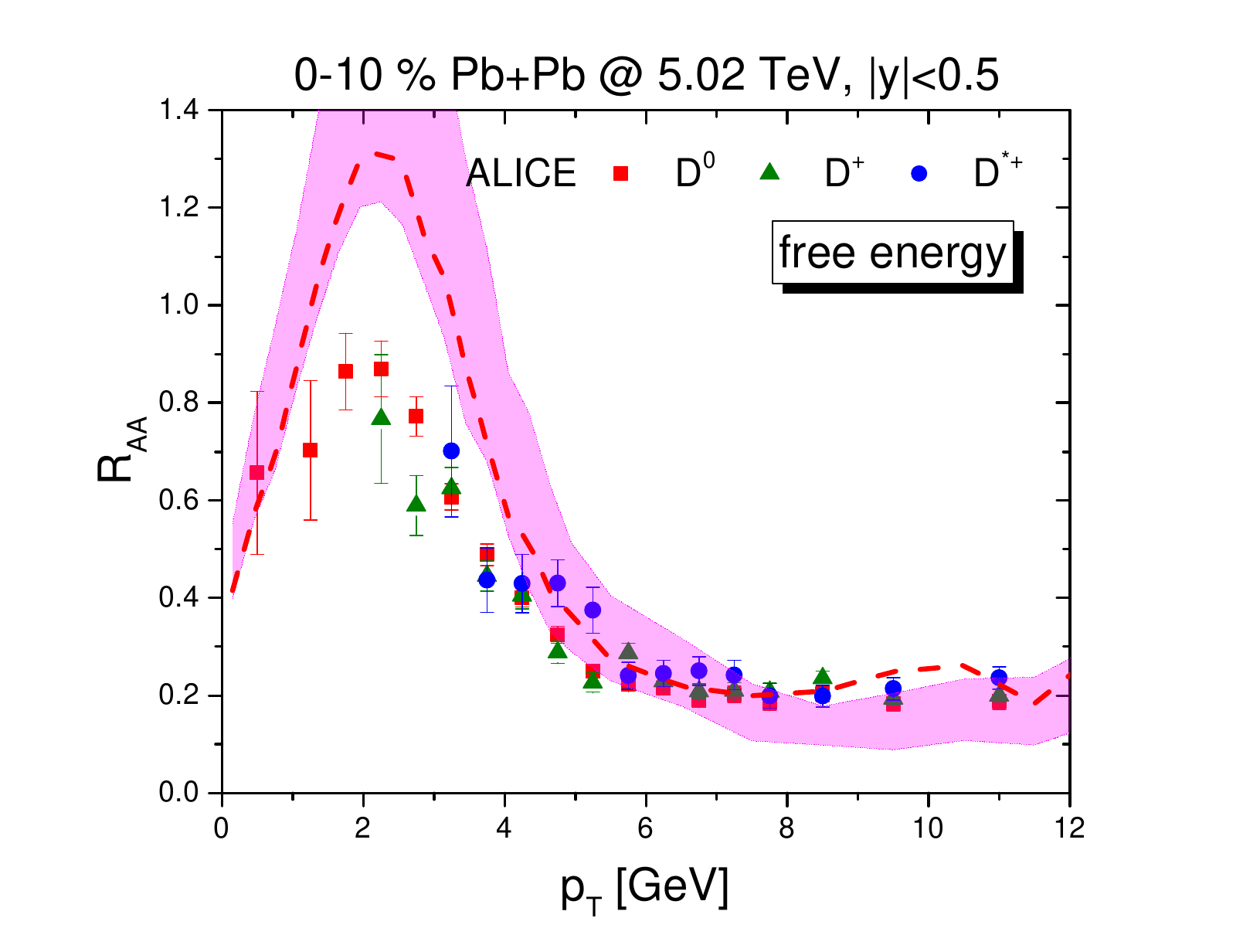}
\includegraphics[width=8.5 cm]{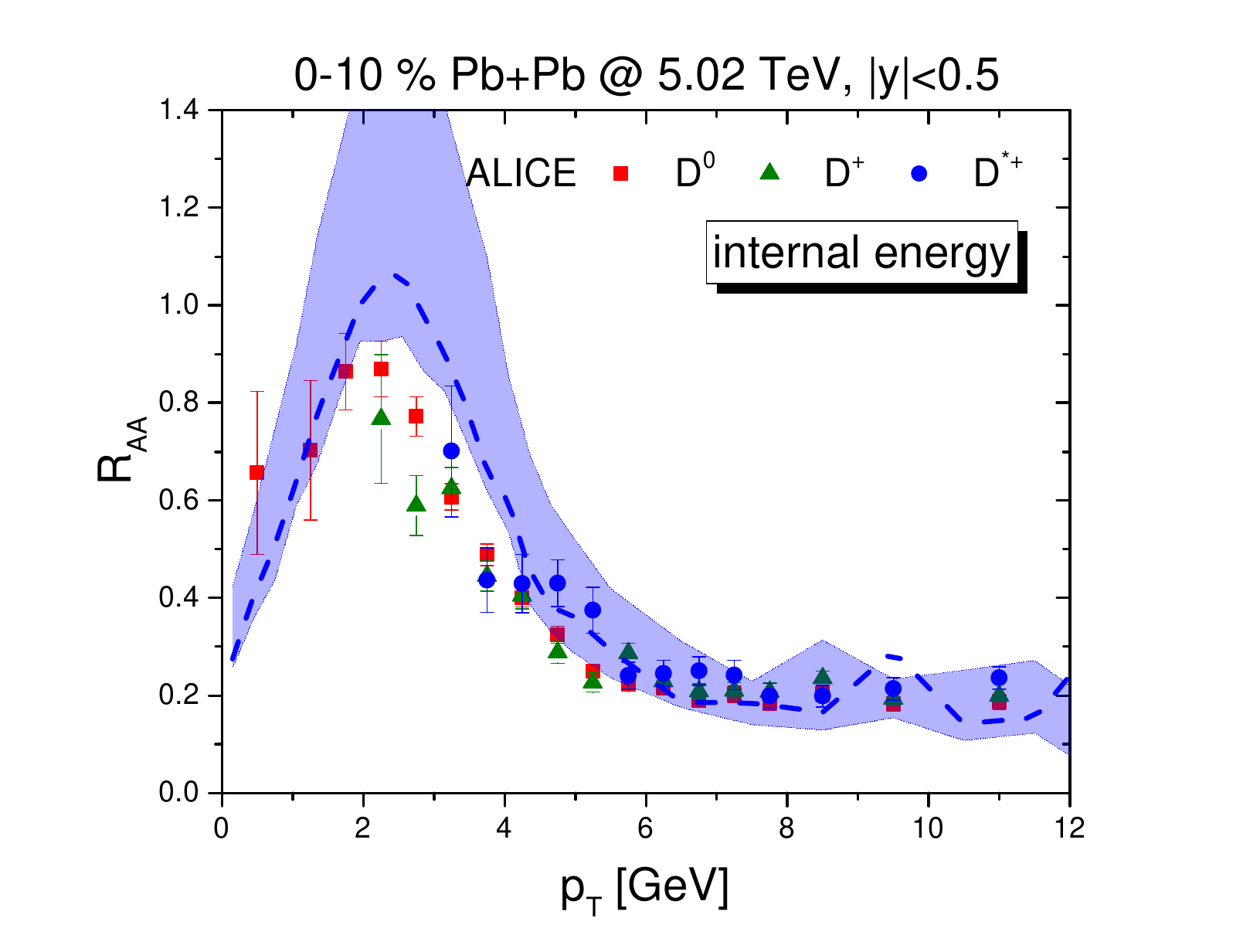}
\includegraphics[width=8.5 cm]{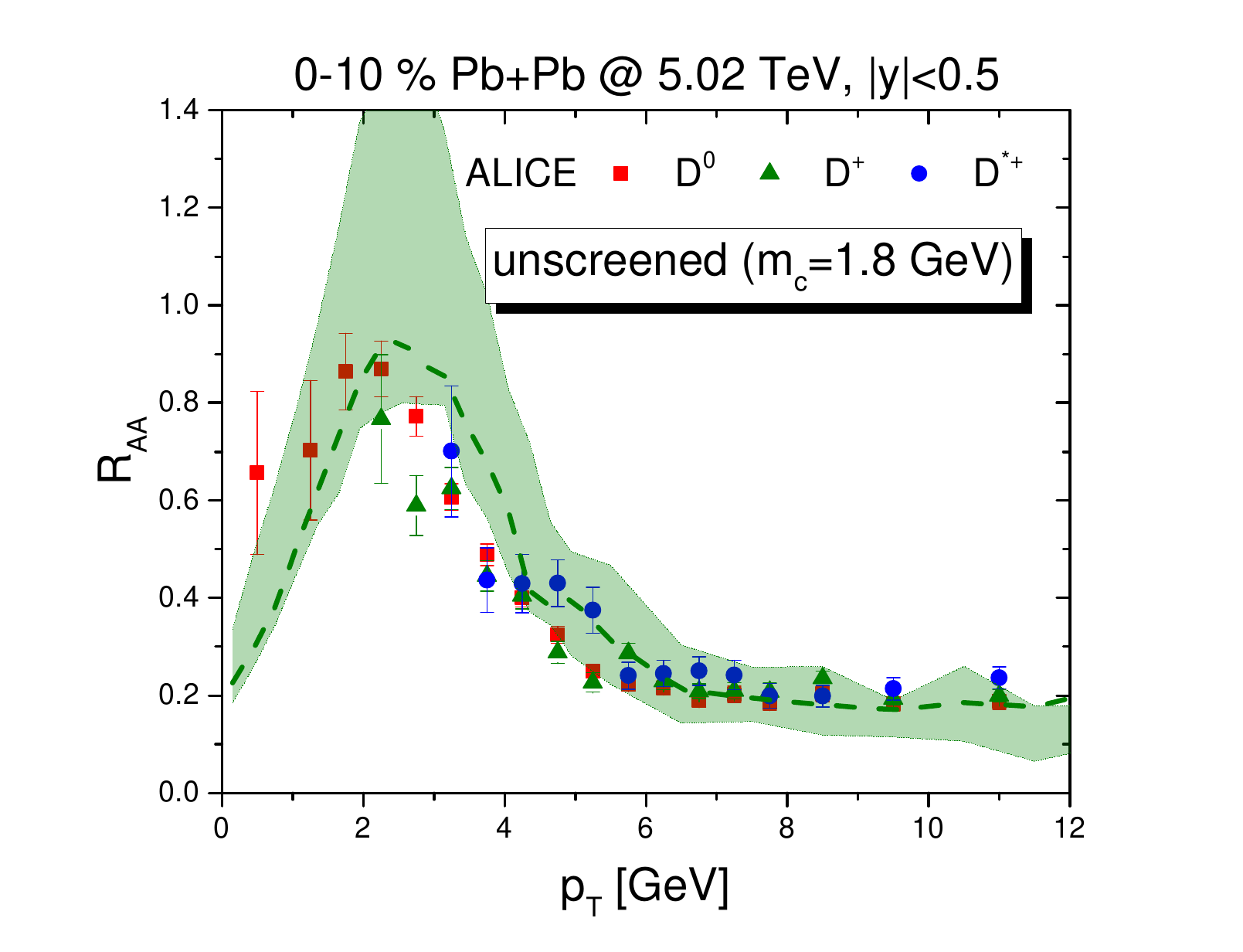}
    \caption{$R_{\rm AA}$ of $D$ meson as a function of $p_T$ in 0-10 \% central Pb+Pb collisions at $\sqrt{s_{\rm NN}}=$ 5.02 TeV for three different heavy quark potential scenarios from the PHSD. The results are compared with the experimental data from the ALICE Collaboration~\cite{ALICE:2021rxa}. The dotted lines indicate the results with the optimized shadowing effects, while the colored bands use two error sets which are most sensitive to charm production in the EPS09~\cite{Eskola:2009uj}.}
    \label{raa-D}
\end{figure}

Figure~\ref{raa-D} shows $R_{\rm AA}$ of $D$ meson as a function of $p_T$ in 0-10 \% central Pb+Pb collisions at $\sqrt{s_{\rm NN}}=$ 5.02 TeV, including thermal charm production for the three different scenarios of effective charm quark mass, and a comparison with experimental data from the ALICE Collaboration~\cite{ALICE:2021rxa}. Since thermal charm mostly contributes to the low $p_T$ region, it is shown only up to $p_T = $12 GeV. In fact, the $R_{\rm AA}$ of $D$ meson above $p_T=6$ GeV barely changes by thermal production.

As mentioned above, the parton distribution function is modified in heavy nucleus, which brings about the (anti)shadowing effects depending on the longitudinal momentum fraction $x$ of parton. The shadowing effects suppress charm production in mid-rapidity and small $p_T$. We use the EPS09 package for the (anti)shadowing effects~\cite{Eskola:2009uj}, which parameterizes the modification of parton distribution function from numerous experimental data in pA and AA collisions.

The dashed lines in Fig.~\ref{raa-D} show the results from the optimized parameter set. There are 15 parameters in the EPS09, and the colored bands are obtained from two error sets (pset=16,17), which are most sensitive to charm production in central Pb+Pb collisions at $\sqrt{s_{\rm NN}} = 5.02$ TeV. One can see that effective charm quark mass from the free energy potential overestimates the experimental data even when including error bars from the shadowing effects.
The results from the internal energy is better than those from the free energy, but they still overestimate charm production, because effective charm quark mass decreases with temperature as shown in Fig.~\ref{charm-mass}, and thermal production is most active at high temperature. 

We obtain the best results from the constant effective charm quark mass, because it suppresses charm thermal production. In this respect the unscreened heavy quark potential in QGP seems consistent with the experimental data on charm production in heavy-ion collisions at LHC.

\section{conclusion}
\label{conclusion}

Since long time ago the thermal production of charm has been anticipated in relativistic heavy-ion collisions especially at LHC. However, the experimental data from LHC doees not show any evidence for the significant thermal production of charm. One possible answer for this is that charm quark gains mass in QGP. This is related to the heavy quark potential in QGP, considering that the energy of heavy quark pair completely separated from each other is equivalent to the mass of two dressed heavy quarks.

In this study we have tested three different heavy quark potentials: the free energy, internal energy of heavy quark pair in QGP, as the screened potentials, and the unscreened potential from the recent lattice QCD study. The cross sections for thermal charm production from quark-antiquark annihilation and two gluon fusion are calculated in the DQPM, and their reverse reactions are also considered for consistency. They are based on the success of heavy quark scattering in the DQPM for the spatial diffusion coefficient of heavy quark from lattice QCD and experimental data on $R_{\rm AA}$ of $D$ meson in heavy-ion collisions. For the simulation of heavy-ion collisions we have used the PHSD, which has been proved successful for the production and dynamics of various kinds of particles, including heavy flavors, in heavy-ion collisions.

Our results show that the free energy potential overestimates charm production, even when considering the uncertainties from the shadowing effects, which are realized by the EPS09 package. The internal energy is better but still overestimates because effective charm quark mass decreases with temperature, while thermal production mostly occurs at high temperatures. We obtain the most consistent results with the experimental data from the unscreened heavy quark potential in QGP, since effective charm quark mass does not decrease with temperature. 

Although our study is relevant to the heavy quark potential only at large distances, it provides valuable information for understanding the heavy quark potential in a thermal medium, which is critical for studying quarkonium in heavy-ion collisions.

\section*{Acknowledgements}
The authors acknowledge inspiring discussions with E. Bratkovskaya. Furthermore, we acknowledge support by the Deutsche Forschungsgemeinschaft (DFG, German Research Foundation) through the grant CRC-TR 211 'Strong-interaction matter under extreme conditions' - Project number 315477589 - TRR 211. The computational resources have been provided by the LOEWE-Center for Scientific Computing and the "Green Cube" at GSI, Darmstadt and by the Center for Scientific Computing (CSC) of the Goethe University, Frankfurt.

\bibliography{main}

\end{document}